# Metasurfaces with Interleaved Electric and Magnetic Resonances for Broadband Arbitrary Group Delay in Reflection

**O. Tsilipakos** [1], Th. Koschny [2], C. M. Soukoulis [1,2]

[1] Institute of Electronic Structure and Laser, FORTH, 71110, Heraklion, Crete, Greece
[2] Ames Laboratory and Department of Physics and Astronomy, Iowa State University, Ames, Iowa, 50011, USA
otsilipakos@iesl.forth.gr

*Abstract* – **Metasurfaces impart phase discontinuities on impinging electromagnetic waves that are typically limited to 0−2π. Here, we show that they can break free from this limitation and supply arbitrarily-large phase modulation over ultra-wide bandwidths. This is achieved by implementing multiple, properly-arranged resonances in the electric and magnetic sheet admittivities. We demonstrate metasurfaces that can perfectly reflect a broadband pulse imparting a prescribed group delay without distorting the pulse shape, opening new possibilities for temporal control and dispersion engineering across deeply subwavelength physical scales.**

## I. INTRODUCTION

Metasurfaces, i.e., ultra-thin periodic configurations of subwavelength resonant meta-atoms, have attracted considerable interest in recent years for an abundance of applications. Providing both electric and magnetic response (i.e., combining electrically and magnetically polarizable meta-atoms) allows for unidirectional scattering —a direct generalization of the Kerker conditions. In addition, the combination of electrically- and magnetically-resonant behavior can extend the underlying phase modulation to 2π. High transmission with an underlying 2π phase shift can be achieved with a pair of matched electric and magnetic resonances and allows —by Huygens' principle— for full control over the wavefront [1]. However, the delay-bandwidth product is still limited, restricted by the maximum 2π shift obtained over the narrow bandwidth of the matched resonance pair, thus limiting the potential of metasurfaces for dispersion engineering applications.

In this work, we demonstrate that metasurfaces can overcome this longstanding limitation on the imparted phase discontinuity. This is achieved by abandoning the singly-resonant nature of conventional metasurfaces in favor of the collective response of *multiple, properly-arranged* electric and magnetic resonances in the effective sheet admittivities. We focus on operation in reflection and specify the required admittivities for achieving arbitrarily-broadband uniform response with an underlying phase modulation greatly exceeding 2π. As an example capturing the rich possibilities of the proposed approach, we construct tunable group delay metasurfaces by requiring an exactly linear reflection phase; this way broadband pulses can experience the desired group delay with zero pulse distortion. We provide realistic implementations with few Lorentzian resonances accommodating broadband signals and assess the performance by studying the available spectral and angular bandwidth.

## II. METASURFACES WITH INTERLEAVED RESONANCES FOR ACCUMULATIVE REFLECTION PHASE

In three-dimensional structures, providing a monotonic reflection phase greatly exceeding 2π is straightforward. Consider for example the grounded dielectric slab (also known as Gires-Tournois etalon) depicted in Fig. 1(a), exhibiting multi-resonant behavior by relying on phase accumulation in the bulk. In metasurfaces, on the other hand, phase delay can only be provided by effective material resonances (i.e., those of the constituent meta-atoms). Our first goal is to determine the proper way of arranging multiple resonances so that the respective shifts combine constructively adding up to an accumulative monotonic reflection phase.

Consider a metasurface described by macroscopic, homogenized electric and magnetic resonant surface admittivities $\sigma_{se}$ and $\sigma_{sm}$. The equations relating the surface admittivities with reflection and transmission coefficients are $r(\omega,\theta) = (-\tilde{\sigma}_e + \tilde{\sigma}_m)/(1 + \tilde{\sigma}_e\tilde{\sigma}_m + \tilde{\sigma}_e + \tilde{\sigma}_m)$ and $t(\omega,\theta) = (1 - \tilde{\sigma}_e\tilde{\sigma}_m)/(1 + \tilde{\sigma}_e\tilde{\sigma}_m + \tilde{\sigma}_e + \tilde{\sigma}_m)$ [2], where we have defined effective sheet admittivities $\tilde{\sigma}_{se}(\omega,\theta) = \zeta\sigma_{se}/2$ and $\tilde{\sigma}_{sm}(\omega,\theta) = \sigma_{sm}/(2\zeta)$ with $\zeta^{\text{TE}}(\theta) = \omega\mu/k_\perp = \eta\sec\theta$ or $\zeta^{\text{TM}}(\theta) = k_\perp/(\omega\varepsilon) = \eta\cos\theta$ depending on the polarization.

Typically, metasurfaces feature a singly-resonant behavior in the electric admittivity, resulting in a narrow high-refection band with an underlying π phase shift. Increasing the reflection phase span and bandwidth by multiple *single-nature* resonances is not possible, as shown in Fig. 1(b). At the point where the respective susceptivity (imaginary part of admittivity) contributions compensate each other, the polarization shifts from anti-phase to in-phase and the reflection phase from 3π/2 to π/2. Reflection phase monotonicity can be achieved by interleaving electric and magnetic resonances, Fig. 1(b). In this case, the electric susceptivity zero-crossing is masked by the magnetic resonance and the reflection phase shows up monotonic. Notably, Fig. 1 illustrates that ultra-thin metasurfaces can emulate the scattering properties of thick bulk structures by replacing the phase accumulation in the bulk with a series of effective material resonances implemented on the surface itself. Note that the transmission phase in both cases of Fig. 1(b) is non-monotonic. Transmission phase spans greatly exceeding 2π can be achieved by concatenating multiple sets of matched resonances [3].

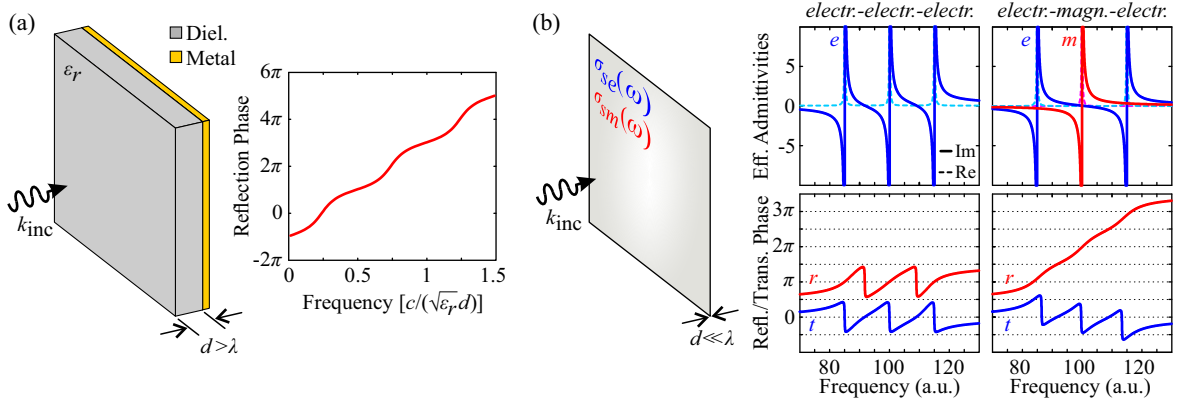

Fig. 1. (a) Grounded dielectric slab: 3D resonant structure with monotonic accumulative reflection phase relying on phase accumulation in the bulk. (b) Multi-resonant 2D metasurface: Interleaving electric and magnetic resonances results in a monotonic accumulative phase profile.

### III. ARBITRARILY-BROADBAND FLAT REFLECTION WITH ZERO GROUP DELAY DISPERSION

Gaining ample control over the reflection phase can be exploited for performing dispersion engineering across deeply subwavelength physical scales. As an example, we focus on metasurfaces that can delay arbitrarily-broadband signals without pulse distortion. Mathematically, we demand $r(\omega) = -\mathcal{A}\exp(i\tau_0\omega)$ and $t(\omega) = 0$, where $\tau_0$ is the desired group delay and $\mathcal{A}$ allows for some absorption in the metasurface. The effective admittivities that strictly satisfy these scattering coefficients are $\tilde{\sigma}_{se} = \tilde{\sigma}_{sm}^{-1} = (1+\mathcal{A}e^{i\tau_0\omega})/(1-\mathcal{A}e^{i\tau_0\omega})$ [2]. Approximating them with physically-realizable Lorentzian resonances we end up with the recipe [2]:

$$\tilde{\sigma}_{se} = \frac{i\kappa_e/2}{\omega + i\Gamma_e^{\text{cor}}/2} + \sum_{k=1}^{+\infty}\frac{i\kappa_e\omega}{\omega^2 - \omega_{e,k}^2 + i\Gamma_e\omega}, \qquad \tilde{\sigma}_{sm} = \sum_{k=1}^{+\infty}\frac{i\kappa_m\omega}{\omega^2 - \omega_{m,k}^2 + i\Gamma_m\omega}, \tag{1}$$

where $\omega_{e,k} = [(2k\pi)^2 + \ln^2(\mathcal{A})]^{1/2}/\tau_0$ ($k = 1,2,\ldots$), $\omega_{m,k} = \{[(2k+1)\pi]^2 + \ln^2(\mathcal{A})\}^{1/2}/\tau_0$ ($k = 0,1,2,\ldots$), $\kappa_e = \kappa_m = 4/\tau_0$, $\Gamma_e = \Gamma_m = 2|\ln(\mathcal{A})|/\tau_0$, and $\Gamma_e^{\text{cor}} = [2(1-\mathcal{A})/(\mathcal{A}|\ln\mathcal{A}|)]\Gamma_e$. The physical recipe of Eq. (1) is plotted in Fig. 2(a) for $\mathcal{A} = 0.9$.

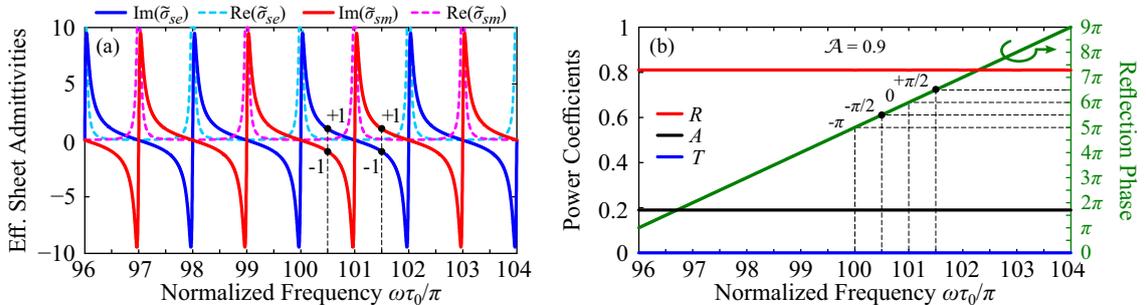

Fig. 2. (a) Lorentzian recipe of Eq. (1) for $\mathcal{A} = 0.9$. (b) Metasurface response: Uniform reflection and linear phase.

The interleaved resonances guarantee the monotonicity of the reflection phase and the specific frequency spacing, strength, and damping are required for providing uniform reflection and an exactly linear phase [Fig. 2(b)]. Note that the linear phase relies on characteristic points with $\mathrm{Arg}(r) = \mp\pi/2$ (corresponding to balanced, anti-phase electric and magnetic contributions) occurring at the midpoints between resonances. Then, the evolution from PEC- to PMC-style reflection happens symmetrically leading to an exactly linear phase profile.

## IV. Practical Implementation With Few Lorentzian Resonances

In practical implementations the Lorentzian sums of Eq. (1) will be truncated to a limited number of terms. In Fig. 3 we keep three electric and four magnetic resonances; their positions are marked with circles and crosses, respectively. Due to truncation, the reflection amplitude and group delay deviate slightly from the prescribed [Fig. 3(a) and (b)]. However, an incident pulse with a spectral bandwidth of 4 that is accommodated inside the reflection band (FWHM~6) experiences the prescribed group delay without broadening or distortion [Fig. 3(c)]. Figure 3(d) illustrates that the proposed metasurfaces are characterized by ample angular bandwidth. When impinging with an incidence angle $\theta_{\mathrm{act}}$ different than the prescribed $\theta_{\mathrm{pre}}$, we get pulse replicas in the temporal domain [2]. However, their amplitude compared to the main pulse is low even for very large angle deviations.

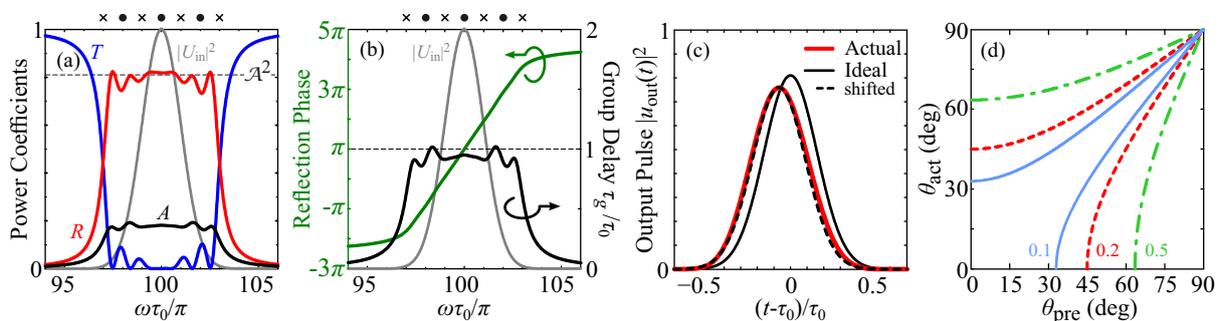

Fig. 3. Response of metasurface with three electric (●) and four magnetic (×) resonances: (a) Scattering coefficients and (b) reflection phase and group delay. The input pulse spectrum, $U_{\mathrm{in}}(\omega)$, is included. (c) Output pulse with negligible distortion. (d) Fractional amplitude of pulse replicas when the incidence angle $\theta_{\mathrm{act}}$ deviates from the prescribed $\theta_{\mathrm{pre}}$.

## V. Conclusion

We have demonstrated that multi-resonant metasurfaces with interleaved electric and magnetic resonances can act as ultra-thin, broadband, true time delay components in reflection, highlighting the potential of performing dispersion engineering across subwavelength scales. Regarding physical implementation, one choice for the meta-atoms is the cut-wire pair which supports closely spaced electric and magnetic resonances [4]. Arranging multiple cut-wire pairs inside the unit cell in an appropriate topology can offer a route to physically implementing the proposed concept.

## Acknowledgement

Work at FORTH was supported by the European Research Council under ERC Advanced Grant no. 320081 (PHOTOMETA) and the European Union's Horizon 2020 Future Emerging Technologies call (FETOPEN-RIA) under grant agreement no. 736876 (VISORSURF). Work at Ames Laboratory was supported by the Department of Energy (Division of Materials Sciences and Engineering) under contract no. DE-AC02-07CH11358.